\documentclass[a4paper,11pt]{article}
\usepackage{jcappub} 
\usepackage{lineno, bm,graphicx,dcolumn,epstopdf,epsf, latexsym,mathbbol, amssymb,amsmath,color,slashed, mathrsfs,mathcomp,simplewick}

 \newcommand{\bq}{\begin{equation}}
 \newcommand{\eq}{\end{equation}}
 \newcommand{\bqn}{\begin{eqnarray}}
 \newcommand{\eqn}{\end{eqnarray}}
 \newcommand{\ban}{\begin{align}}
 \newcommand{\ean}{\end{align}}
  \newcommand{\nb}{\nonumber}
 \newcommand{\lb}{\label}

\arxivnumber{2403.05841} 

\title{\boldmath Power spectra and circular polarization of primordial gravitational waves with parity and Lorentz violations}

\author[a, b]{Tian-Chen Li,}
\author[a, b, 1]{Tao Zhu \note{Corresponding author},}
\author[c, d]{Wen Zhao,}
\author[e]{and Anzhong Wang}
\affiliation[a]{Institute for Theoretical Physics \& Cosmology, Zhejiang University of Technology, Hangzhou, 310023, China}
\affiliation[b]{United Center for Gravitational Wave Physics,  Zhejiang University of Technology, Hangzhou, 310023, China}
\affiliation[c]{Department of Astronomy, University of Science and Technology of China, Hefei, Anhui 230026, China}
\affiliation[d]{School of Astronomy and Space Science, University of Science and Technology of China, Hefei 230026, China}
\affiliation[e]{GCAP-CASPER, Physics Department, Baylor University, Waco, TX 76798-7316, USA}

\emailAdd{zhut05@zjut.edu.cn}

\abstract{
The violations of parity and Lorentz symmetries in gravity can change the propagating properties of gravitational waves (GWs) in the cosmological background, which can arise from a large number of parity- and Lorentz-violating theories. In this paper, through a systematic parametrization for characterizing possible derivations from the standard GW propagation in general relativity, we study both the parity- and Lorentz-violating effects on the power spectra and the polarization of the primordial gravitational waves (PGWs) during the slow-roll inflation. To this end, we calculate explicitly the power spectrum and the corresponding circular polarization of the PGWs analytically by using the uniform asymptotic approximation. It is shown that the new contributions to power spectra contain two parts, one from the parity-violating terms and the other from the Lorentz-violating terms. While the Lorentz-violating terms can only affect the overall amplitudes of PGWs, the parity-violating terms induce nonzero circular polarization of PGWs, i.e., the left-hand and right-hand polarization modes of GWs have different amplitudes. }

\begin{document}
\maketitle
\flushbottom

\section{Introduction}
\renewcommand{\theequation}{1.\arabic{equation}}\setcounter{equation}{0}
\label{sec:intro}

The inflationary paradigm offers a dominant framework for understanding the generations of both the primordial density perturbations and primordial gravitational waves (PGWs) \cite{Guth:1980zm, Starobinsky:1980te, Sato:1980yn, Baumann:2009ds}. The former provides primordial seeds for the formation of the large-scale structure (LSS) in the Universe as observed today, and is also responsible for the temperature anisotropies detected in the cosmic microwave background (CMB) by numerous experiments, including WMAP \cite{WMAP:2012nax, Larson:2010gs} and PLANCK \cite{Planck:2018vyg, Planck:2018jri}. On the other hand, PGWs can imprint distinct signatures in the CMB spectra \cite{Krauss:2010ty, Garcia-Bellido:2010uka, Bock:2006yf, Seljak:1996gy, Kamionkowski:1996zd} and influence the galaxy power spectrum \cite{Schmidt:2012nw, Schmidt:2013gwa, Dodelson:2003bv, Dodelson:2010qu, Schmidt:2012ne, Chisari:2014xia, Biagetti:2020lpx}. In the CMB, PGWs are expected to generate TT, EE, BB, and TE spectra, while the TB and EB spectra should be absent if parity symmetry in gravity is preserved \cite{Krauss:2010ty, Garcia-Bellido:2010uka, Bock:2006yf, Seljak:1996gy, Kamionkowski:1996zd}. This makes their observation a crucial goal for forthcoming CMB experiments \cite{SimonsObservatory:2018koc, Hazumi:2019lys, Li:2017drr, CMB-S4:2016ple, shamik}. Similarly, in the galaxy power spectrum, PGWs leave discernible effects in the B-mode polarization, whereas the E-B correlation is expected to vanish when parity conservation holds \cite{Schmidt:2012nw, Schmidt:2013gwa, Dodelson:2003bv, Dodelson:2010qu, Schmidt:2012ne, Chisari:2014xia, Biagetti:2020lpx}. As such, upcoming galaxy surveys may offer critical insights into the physics underlying PGWs \cite{Biagetti:2020lpx, Amendola:2016saw, LSST:2008ijt}.

Most inflation models that produce PGWs use the theory of general relativity (GR) as the theory of gravity. In GR, parity and Lorentz symmetries are two fundamental symmetries. However, possible violations of these two symmetries may arise in theories that try to unify quantum physics with gravity. Because of this, various modified theories of gravity have been proposed to explore the nature of parity and Lorentz violations in gravity, to mention a few, including the Chern-Simons modified gravity \cite{Jackiw:2003pm, Alexander:2009tp}, the symmetric teleparallel equivalence of GR theory \cite{Conroy:2019ibo, Li:2022vtn}, Horava-Lifshitz theories of quantum gravity \cite{Horava:2009uw, Blas:2009qj, Zhu:2011yu, Zhu:2011xe, Takahashi:2009wc, Colombo:2014lta, Wang:2017brl}, chiral scalar-tensor theory \cite{Crisostomi:2017ugk}, the Nieh-Yan teleparallel modified gravity \cite{Li:2020xjt, Li:2021wij, Wu:2021ndf}, parity-violating scalar-tensor theory in teleparallel gravity \cite{Rao:2023doc}, parity violation induced by couplings between dual Reimann tensor and Kalb-Ramond two-form field \cite{Manton:2024hyc}, the linearized gravity in standard model extension \cite{Mewes:2019dhj, Kostelecky:2017zob}, Einstein-\AE{}ther theories \cite{Jacobson:2000xp, Eling:2004dk, Jacobson:2007veq, Oost:2018oww, Zhang:2019iim, Tsujikawa:2021typ, Zhang:2023kzs, Zhang:2022fbz}, and the spatial covariant gravities \cite{Gao:2019liu, Gao:2020qxy, Gao:2020yzr, Gao:2019twq, Yu:2024drx, Joshi:2021azw}, etc. 

The violations of parity and Lorentz symmetries in gravity could induce possible derivations from the standard propagation properties of GWs in GR, the two independent GW polarizations propagate at the speed of light with an amplitude damping rate as the inverse of the luminosity distance of the GW sources. Different mechanisms of parity and Lorentz violations may induce different effects in GW propagations. For example, parity violation in gravity in general can induce an asymmetry of the propagation speed and amplitude damping between left- and right-hand polarizations of a GW, which leads to the velocity and amplitude birefringence, respectively. On the other hand, the Lorentz violation can result in two distinct effects on the propagation of GWs. First, with Lorentz violation, the conventional linear dispersion relation of GWs can be modified into a nonlinear one, which in turn changes the phase velocities of GWs at different frequencies. Second, Lorentz violation can introduce frequency-dependent friction into the propagation equation of GWs, resulting in varying damping rates for GWs of different frequencies during their propagation. 

Thus, it is expected that the above-mentioned new effects could induce some distinguishable signatures in the power spectrum of PGWs. Such considerations have attracted a great deal of attention lately and several phenomenological implications of both parity and Lorentz violations on the power spectra of PGWs have already been investigated in several specific parity- and Lorentz-violating theories of gravity, see refs. \cite{Li:2022xww, Qiao:2022mln, Zhu:2022dfq, Fu:2020tlw, Qiao:2019hkz, Ding:2019nwu, Qiao:2018dpp, Zhu:2016srz, Kawai:2017kqt} and references therein. On the other hand, considering that there are a large number of parity- and Lorentz-violating gravities, it is challenging to study their observational effects on the PGWs systematically, instead of studying each specific theory separately. Recently, a systematic parametric framework for characterizing possible derivations of GW propagation in GR was constructed in refs. \cite{Zhao:2019xmm, Zhu:2023wci}. This parameterization provides a general framework for studying the GW propagation of possible modifications caused by various modified gravitational theories \cite{Zhao:2019xmm, Zhu:2023wci}. This parameterization has also been used to study the effect of parity/Lorentz violation on the modified GW waveforms of the binary inspiral system and their constraints from the data of compact binary merging events detected by LIGO-Virgo-KAGRA collaboration \cite{Zhu:2023wci}. For other alternative parametrized frameworks, see refs.~\cite{Nishizawa:2017nef, Ezquiaga:2021ler, Tahura:2018zuq, Saltas:2014dha}. 

In this paper, through the systematic parametrization for characterizing possible derivations from GW propagation in GR constructed in refs.~\cite{Zhao:2019xmm, Zhu:2023wci}, we investigate both the parity- and Lorentz-violating effects on the power spectra and the polarization of PGWs. For this purpose, we employ the third-order uniform asymptotic approximation to calculate explicitly the power spectrum and the corresponding circular polarization of the PGWs analytically. We also compare our approximate results with those special cases with exact results. In addition, the observational implications of the parity- and Lorentz-violating effects on PGWs are also briefly discussed. 

The structure of this paper is organized as follows. In Section II, we introduce the parameterization for characterizing the possible derivations from GW propagation in GR. With this parameterization, we then derive the master equation that describes the propagation of GWs during inflation in Sec. III. In Sec. IV, by using the third-order uniform asymptotic approximation, we calculate explicitly the power spectra and the polarization of PGWs during the slow-roll inflation. We also compare our approximate results with those special cases which have exact results in Sec. V.  Our main conclusions and outlook are summarized in Sec. VI.

Throughout this paper, the metric convention is choosen as $(-,+,+,+)$, and greek indices $(\mu, \nu, \cdots)$ run over $0$, $1$, $2$, $3$ and latin indices $(i, j, k, \cdots)$ run over $1$, $2$, $3$. 

\section{Parameterization for characterizing possible derivations of GW propagation}
\renewcommand{\theequation}{2.\arabic{equation}}\setcounter{equation}{0}

In this section, we provide an introduction of a universal parameterization, for characterizing the possible derivations from GW propagation in GR \cite{Zhao:2019xmm, Zhu:2023wci}. 
To start, let us consider GWs propagating on a homogeneous and isotropic background,
\bqn
ds^2=a^2(\tau)[-d\tau^2 +(\delta_{ij}+h_{ij})dx^i dx^j],
\eqn
where $a(\tau)$ is the conformal scale factor, and we set its resent value as $a_0=1$ in this paper. $\tau$ represents the conformal time which relates to the cosmic time by $dt=ad\tau$. $x^i$ is the comoving coordinates. The GWs are represented by $h_{ij}$ which are transverse and traceless, i.e.,
\bqn
\delta^{ij} h_{ij} =0, \;{\rm and}\; \partial_ih^{ij}=0.
\eqn
It is convenient to expand $h_{ij}$ over spatial Fourier harmonics,
\bqn
h_{ij}(\tau, x^i) = \sum_{A=R,L}\int \frac{d^3k}{(2\pi)^3} h_A(\tau, k^i) e^{i k_ix^i} e_{ij}^A(k^i),
\eqn
where $e_{ij}^A$ is the circular polarization tensor and obeys the following rules
\begin{eqnarray}
    \epsilon^{ijk}n_i e^A_{kl} = i \rho_A e_l^{j A}
\end{eqnarray}
with $\rho_{\rm R}=1$ and $\rho_{\rm L}=-1$. Here the unite vector $n^i\equiv k^i/k$ with $k= (k_i k^i)^{1/2}$ which satisfies $n^i e_{ij}^A =0$. 

To study the Lorentz and parity-violating effects in the propagation of GWs, let us first write the modified propagation equations of the two GW modes in the following parametrized form \cite{Zhao:2019xmm, Zhu:2023wci}
\begin{equation}
\label{h1}
h''_A+(2+\bar{\nu}+\nu_A)\mathcal{H}h'_A+(1+\bar{\mu}+\mu_A)k^2 h_A=0,
\end{equation}
where a prime denotes the derivation with respect to the conformal time $\tau$ and $\mathcal{H} = a'/a$. Then the new effects caused by different modified theories are characterized by four parameters, $\bar{\nu}$, $\bar{\mu}$, $\nu_A$, and $\mu_A$. These effects can be divided into the following three categories:  1) the Lorentz-violating effects induced by frequency-dependent $\bar{\nu}$ and $\bar{\mu}$, including the frequency-dependent damping and nonlinear dispersion relation of GWs; 2) the parity-violating effects induced by frequency-dependent $\nu_A$ and $\mu_A$, including the amplitude and velocity birefringences of GWs; and 3) the frequency-independent effects induced by frequency-independent $\bar{\nu}$ and $\bar{\mu}$, including modifications of the GW friction and speed. 

The parameters $\bar{\mu}$ and $\bar \nu$, when they are frequency-independent, correspond to the modifications to the GW speed and friction, respectively. These two modifications can arise from a broad class of modified gravities that are generally irrelevant for parity and Lorentz violations, see Table I of ref. \cite{Zhu:2023wci} for the relevant modified theories. Some of Lorentz-violating theories, in which the Lorentz violations can be induced by the presence of the background fields including the Einstein-\AE{}ther theory \cite{Foster:2006az, Oost:2018tcv},  the bumblebee gravity \cite{Liang:2022hxd}, and Lorentz violation with an antisymmetric tensor \cite{Altschul:2009ae} and a Kalb-Ramond field \cite{Manton:2024hyc}, can induce modifications to GW speed and friction as well. In the slow-roll inflation phase, these two modifications lead to an effective sound speed and mass squared terms in the equation of motion of the PGWs. By treating all the slow-roll quantities in the equation as constants, the corresponding analytical solution of the PGWs can be solved exactly. Their effects on the power spectra of PGWs have been well-understood and studied extensively in a broad class of modified gravities. Therefore, this paper only focuses on the PGWs with new effects when the four parameters $\bar{\nu}$, $\bar{\mu}$, $\nu_A$, and $\mu_A$ are frequency-dependent. 

The parameters $\nu_A$ and $\mu_A$ correspond to the gravitational parity-violating effects. The parameter $\nu_A$ induces amplitude birefringence, leading to different damping rates of left- and right-hand circular polarizations of GWs, while the parameter $\mu_A$ induces velocity birefringence, leading to different velocities of left- and right-hand circular polarizations of GWs. For both $\nu_A$ and $\mu_A$ are frequency-dependent, they can be parametrized as \cite{Zhao:2019xmm, Zhu:2023wci}
\bqn
\mathcal{H}\nu_{A}&=&\left[\rho_{A}\alpha_{\nu}(\tau)\left(\frac{k}{aM_{\rm{PV}}}\right)^{\beta_{\nu}}\right]_{, \tau}, \\
\mu_{A}&=&\rho_{A}\alpha_{\mu}(\tau)\left(\frac{k}{aM_{\rm{PV}}}\right)^{\beta_{\mu}},
\eqn
where $\alpha_{\nu}$, $\alpha_{\mu}$ are arbitrary functions of time and $\beta_{\nu}$, $\beta_{\mu}$ are arbitrary numbers, $M_{\rm{PV}}$ denotes the energy scale of the parity violation, and ``$,\tau$" represents the derivative with respect to $\tau$. In principle, the frequency dependencies of ${\cal H} \nu_A$ and $\nu_A$ may encompass multiple $k$-terms characterized by different integers $\beta_{\nu}$ and $\beta_\mu$, as well as distinct time-dependent coefficients, as parameterized in \cite{Jenks:2023pmk, Daniel:2024lev}. Indeed, as evidenced by Table I in ref.~\cite{Zhu:2023wci}, certain modified theories can give rise to more than one non-GR parameter with multiple $k$-depencence terms. However, in our parametrization \cite{Zhao:2019xmm, Zhu:2023wci}, we assume that among the various $k$-terms in each non-GR parameter, only one term is dominant. We focus on this dominant term to consider its potential effects on observables generally. It is also fairly straightforward to add up the effects of different $k$-terms if needed.

The violations of Lorentz symmetry or diffeomorphisms can lead to nonzero and frequency-dependent $\bar{\nu}$ and $\bar{\mu}$. The parameter $\bar{\mu}$ induces frequency-dependent friction in the propagation equation of GWs, while $\bar{\mu}$ modifies the conventional linear dispersion relation of GWs to nonlinear ones. For both $\bar{\nu}$ and $\bar{\mu}$ are frequency-dependent, they can be parametrized as \cite{Zhao:2019xmm, Zhu:2023wci}
\bqn
\mathcal{H}\bar{\nu}&=&\left[\alpha_{\bar{\nu}}(\tau)\left(\frac{k}{aM_{\rm{LV}}}\right)^{\beta_{\bar{\nu}}}\right]_{, \tau}, \\
\bar{\mu}&=&\alpha_{\bar{\mu}}(\tau)\left(\frac{k}{aM_{\rm{LV}}}\right)^{\beta_{\bar{\mu}}},
\eqn
where $\alpha_{\bar{\nu}}$, $\alpha_{\bar{\mu}}$ are arbitrary functions of time and $\beta_{\bar{\nu}}$, $\beta_{\bar{\mu}}$ are arbitrary numbers. $M_{\rm{LV}}$ means the energy scale of the Lorentz violation.

The corresponding modified theories with specific forms of the four parameters $\cal{H}\bar{\nu}$, $\bar{\mu}$, ${\cal H}\nu_A$, and $\mu_A$, and their GW frequency dependences are summarized in Table I of ref. \cite{Zhu:2023wci}.

\section{Equation of motion for GWs}
\renewcommand{\theequation}{3.\arabic{equation}}\setcounter{equation}{0}

Let us now study the propagation equations of GWs with the above parametrization for possible parity- and Lorentz-violating effects. For later convenience of calculating the primordial power spectra of GWs, let us introduce a new variable
\bqn
u_A= \frac{1}{2}  z_A h_A,
\eqn
with
\bqn
z_A &=& a \left[1+\alpha_{\bar{\nu}}(\tau)\left(\frac{k}{aM_{\rm{LV}}}\right)^{\beta_{\bar{\nu}}} +\rho_{A}\alpha_{\nu}(\tau)\left(\frac{k}{aM_{\rm{PV}}}\right)^{\beta_{\nu}} \right]^{1/2}.
\eqn
Then the equation of motion (\ref{h1}) can be rewritten as
\bqn
\label{u1}
u''_A+\left(\omega_A^2-\frac{z''_A}{z_A}\right)u_A=0,
\eqn
where
\bqn\label{w1}
\frac{\omega_A^2}{k^2} \equiv 1+\rho_{A}\alpha_{\mu}(\tau)\left(\frac{k}{aM_{\rm{PV}}}\right)^{\beta_{\mu}}+\alpha_{\bar{\mu}}(\tau)\left(\frac{k}{aM_{\rm{LV}}}\right)^{\beta_{\bar{\mu}}}.
\eqn
In the above expression, we expect the derivations from GR to be small. In this way, one can expand $z''_A/z_A$ in terms of non-GR parameters about its GR result, which is
\bqn\label{z1}
\frac{z''_A}{z_A}&\simeq&\left[1-\frac{\beta_{\nu}}{2} \rho_A \alpha_{\nu}\Big(\frac{k}{aM_{\rm{PV}}}\Big)^{\beta_{\nu}}-\frac{\beta_{\bar{\nu}}}{2}\alpha_{\bar{\nu}}\Big(\frac{k}{aM_{\rm{LV}}}\Big)^{\beta_{\bar{\nu}}}\right]\frac{a''}{a} \nb\\
&&+ \frac{1}{2}\left[(\beta_{\nu}^2-\beta_{\nu})\rho_A \alpha_{\nu}\Big(\frac{k}{aM_{\rm{PV}}}\Big)^{\beta_{\nu}} +(\beta_{\bar{\nu}}^2-\beta_{\bar{\nu}})\alpha_{\bar{\nu}}\Big(\frac{k}{aM_{\rm{LV}}}\Big)^{\beta_{\bar{\nu}}}\right]\frac{a'^2}{a^2}\nb\\
&&+\left[(1-\beta_{\nu})\rho_A \alpha'_{\nu}\Big(\frac{k}{aM_{\rm{PV}}}\Big)^{\beta_{\nu}} +(1-\beta_{\bar{\nu}})\alpha'_{\bar{\nu}}\Big(\frac{k}{aM_{\rm{LV}}}\Big)^{\beta_{\bar{\nu}}}\right]\frac{a'}{a}\nb\\
&&+\frac{1}{2}\left[\rho_A \alpha''_{\nu}\Big(\frac{k}{aM_{\rm{PV}}}\Big)^{\beta_{\nu}}+\alpha''_{\bar{\nu}}\Big(\frac{k}{aM_{\rm{LV}}}\Big)^{\beta_{\bar{\nu}}}\right].
\eqn
Note that in the above expansion, we only consider the first-order terms of each coefficient.

In this article, the PGWs are considered during the inflationary stage, with the background evolution varying slowly. Under this condition, all the coefficients $\alpha_{\nu}$, $\alpha_{\bar{\nu}}$, $\alpha_{\mu}$ and $\alpha_{\bar{\mu}}$ can be treated as slowly varying quantities. Then we can expand the modified dispersion relation $\omega_A^2$ in (\ref{w1}) and effective time-dependent mass term $z''_A/z_A$ in (\ref{z1}) in terms of the slow-roll quantities as
\bqn
\frac{\omega^2_A}{k^2}\simeq1+\rho_A\alpha_{\mu}\Big(-\frac{Hk\tau}{M_{\rm{PV}}}\Big)^{\beta_{\mu}}+\alpha_{\bar{\mu}}\Big(-\frac{Hk\tau}{M_{\rm{LV}}}\Big)^{\beta_{\bar{\mu}}},
\eqn
and
\bqn
\frac{z''_A}{z_A}&\simeq&\frac{1}{\tau^2}(2+3\epsilon_1) \nb\\
&& +\frac{\rho_A}{2\tau^2 H^2}\Big(-\frac{Hk\tau}{M_{\rm{PV}}}\Big)^{\beta_{\nu}}\Big[(-3+\beta_{\nu})\beta_{\nu}\alpha_{\nu}H^2  +(3-2\beta_{\nu})\dot{\alpha}_{\nu} H+\ddot{\alpha}_{\nu}\Big]\nb\\
&&+\frac{1}{2\tau^2 H^2}\Big(-\frac{Hk\tau}{M_{\rm{LV}}}\Big)^{\beta_{\bar{\nu}}}\Big[(-3+\beta_{\bar{\nu}})\beta_{\bar{\nu}}\alpha_{\bar{\nu}} H^2 +(3-2\beta_{\bar{\nu}}) \dot{\alpha}_{\bar{\nu}} H+\ddot{\alpha}_{\bar{\nu}}\Big].
\eqn
Here $H=\dot a/a$ denotes the Hubble parameter and a dot represents the derivative with respect to the cosmic time $t$. Note that in deriving the above expressions, we have used the approximate relation
\bqn
a\simeq-\frac{1+\epsilon_1}{\tau H},
\eqn
where the slow-roll parameter $\epsilon_1 \equiv-\dot{H}/H^2$.

With the expressions of $\omega^2_A/k^2$ and $z''_A/z_A$, the equation of motion in (\ref{u1}) can be changed into the form of
\bqn\label{u2}
&& u''_A+\Bigg[1-\frac{v_t^2-\frac{1}{4}}{k^2\tau^2}  \nb\\
&&~~~~~~~~~ + \rho_A d_1 (-k\tau)^{\beta_{\nu}-2}+ d_2 (-k\tau)^{\beta_{\bar{\nu}}-2} + \rho_A d_3 (-k\tau)^{\beta_{\mu}}+ d_4 (-k\tau)^{\beta_{\bar{\mu}}} \Bigg]k^2 u_A=0, \nb\\
\eqn
where
\bqn
v_t^2 &\equiv& \frac{9}{4}+3\epsilon_1,\nb\\
d_1 &\equiv & \Big(\frac{H}{M_{\rm{PV}}}\Big)^{\beta_{\nu}}\frac{\beta_{\nu}(3-\beta_{\nu}) \alpha_{\nu} H^2-(3-2\beta_{\nu})\dot{\alpha}_{\nu} H-\ddot{\alpha}_{\nu}}{2H^2},\nb\\
d_2&=&\Big(\frac{H}{M_{\rm{LV}}} \Big)^{\beta_{\bar{\nu}}}\frac{\beta_{\bar{\nu}}(3-\beta_{\bar{\nu}})\alpha_{\bar{\nu}} H^2-(3-2\beta_{\bar{\nu}})\dot{\alpha}_{\bar{\nu}} H-\ddot{\alpha}_{\bar{\nu}}}{2H^2},\nb\\
d_3&=&\Big(\frac{H}{M_{\rm{PV}}}\Big)^{\beta_{\mu}}\alpha_{\mu},\nb\\
d_4&=&\Big(\frac{H}{M_{\rm{LV}}}\Big)^{\beta_{\bar{\mu}}}\alpha_{\bar{\mu}}.
\eqn
All of these coefficients (i.e. $v_t, d_1, d_2, d_3, d_4$) are slowly varying and dimensionless. It is easy to see that there are no exact and analytical solutions for this equation, even if all the slowly varying quantities are considered as constants. In this paper, we will employ the uniform asymptotic approximation to construct the approximate analytical solutions to the above equations. The uniform asymptotic approximation was developed in a series of papers for better treatment of the second-order ordinary differential equations with turning points and poles, and widely applied in the calculations of primordial spectra of various inflation models \cite{Zhu:2013upa, Qiao:2018dpp, Habib:2002yi, Zhu:2014wfa, Zhu:2013fha, Zhu:2014aea}. This approximation is also used in applications in studying the reheating process \cite{Zhu:2018smk} and quantum mechanics \cite{Li:2019cre}. In the next section, an approximate solution of (\ref{u2}) will be constructed through this approximation. Then the corresponding primordial tensor power spectra with both the parity- and Lorentz-violating effects can be derived.

\section{Power spectra and circular polarization of PGWs in the uniform asymptotic approximation}
\renewcommand{\theequation}{4.\arabic{equation}}\setcounter{equation}{0}

\subsection{General formula for the power spectra}

In this subsection, we present a brief introduction of the general formulas of primordial perturbation spectra in the uniform asymptotic approximation.  Most of the expressions and results used here can be found in \cite{Zhu:2022dfq, Qiao:2019hkz, Qiao:2019hkz, Zhu:2013upa, Zhu:2014aea}.

Firstly, in the uniform asymptotic approximation, we write the equation of motion (\ref{u2}) in the following standard form \cite{olver_asymptotics_1997, Zhu:2013upa, Zhu:2013fha},
\begin{equation}
\label{eqmot}
\frac{d^2 u_A(y)}{dy^2}=\big[g(y)+q(y)\big]u_A(y),
\end{equation}
where $y\equiv -k\tau$ is a dimensionless variable, and
\bqn\label{gq1}
g(y)+q(y)&\equiv&\frac{v_t^2-\frac{1}{4}}{y^2}-1 -\rho_A d_1 y^{\beta_{\nu}-2}- d_2y^{\beta_{\bar{\nu}}-2} -\rho_A d_3 y^{\beta_{\mu}}- d_4 y^{\beta_{\bar{\mu}}}.
\eqn
Normally, the functions $g(y)$ and $q(y)$ given in the above contain poles and turning points. Their properties depend on the values of the four parameters $\beta_\nu$, $\beta_\mu$, $\beta_{\bar \nu}$, and $\beta_{\bar \mu}$. In this paper, we focus on those cases with $\beta_{\nu}, \; \beta_{\bar \nu} \geq 0$ and $\beta_{\mu}, \; \beta_{\bar \mu} \geq -2$. As shown in Table. I of ref.~\cite{Zhu:2023wci}, the values of  $\beta_\nu$, $\beta_\mu$, $\beta_{\bar \nu}$, and $\beta_{\bar \mu}$ with the above conditions include most of the modified gravities with parity and Lorentz violations. For these cases, it is easy to observe that the function $g(y)+q(y)$ at least contains a second-order pole at $y \to 0^+$. It can also have another pole at $y  \to +\infty$ if $\beta_{\nu}, \; \beta_{\bar \nu} >2$ or $\beta_{\mu}, \; \beta_{\bar \mu} >0$.

In the uniform asymptotic approximation, to make the approximate solution valid around the second-order pole (the pole at $y\to 0^+$), one has to ensure that the error control function associated with the approximate solution is convergent. For the equation of motion in (\ref{eqmot}) with $g(y) + q(y)$ given by (\ref{gq1}), it has been proved in \cite{olver_asymptotics_1997, Zhu:2013upa} that to make its error control function to be convergent around the second-order pole at $y = 0^+$, one must select
\bqn
q(y)=-\frac{1}{4y^2}. \lb{qofy}
\eqn
Then $g(y)$ can be written as
\bqn\label{gy}
g(y)&=&\frac{v_t^2}{y^2}-1 -y^{\beta_{\nu}-2}\rho_A d_1 - d_2 y^{\beta_{\bar{\nu}}-2} - \rho_A d_3 y^{\beta_{\mu}} - d_4 y^{\beta_{\bar{\mu}}}.
\eqn

Besides the two poles at $y = 0^+$ and $y = +\infty$, $g(y)$ may have another single zero in the range $y \in (0, +\infty)$, which is called a single turning point of $g(y)$. Since we have $v^2_t \simeq \frac{9}{4} + 3\epsilon_1$, and $d_1 = 0 = d_2 = d_3 = d_4$ in GR, we expect all the new terms with coefficients $d_1$, $d_2$, $d_3$, and $d_4$ can be considered as small corrections. Then with this consideration and solving the equation $g(y) = 0$, we obtain the turning point,
\bqn
y_0 \simeq v_t - \frac{\rho_A d_1 v_t^{\beta_{\nu}-1}+d_2 v_t^{\beta_{\bar{\nu}}-1}+\rho_A d_3 v_t^{\beta_{\mu}+1}+d_4 v_t^{\beta_{\bar{\mu}}+1}}{2}.\nb\\
\eqn

With the above choices of $g(y)$ and $q(y)$, as given in Eqs.~(\ref{qofy}) and (\ref{gy}) respectively, the analytical approximate solution of eq.~(\ref{eqmot}) associated with the single turning point $y_0$ can be constructed in terms of Airy functions. With these analytical and approximate solutions, the general formulas of the power spectra of the PGWs can be written as follows \cite{Zhu:2014wfa},
\bqn
{\cal P}_{\rm T}^{A} &\equiv & \frac{2k^3}{\pi^2} \left|\frac{u_A(y)}{z_A}\right|^2_{y \to 0^+} \nb\\
&\simeq & \frac{k^2}{\pi^2} \frac{- k \tau}{z_A^2 v_t} \exp{\left(2 \int_y^{y_0} \sqrt{g(y')}dy'\right)} \times \left[1+ {\mathscr H}(+\infty) + \frac{{\mathscr H}^2(+\infty)}{2} + \cdots \right]. \label{PW_3}
\eqn
Here we only consider the third-order uniform asymptotic approximation and ``$\cdots$" denotes the contributions beyond the third-order. ${\mathscr H}(+\infty)$ is the error control function which is given by \cite{Zhu:2014wfa}
\bqn\lb{H}
\mathscr{H}(+\infty) &=&\frac{5}{36}\left.\left\{\int_{y_0}^{y}\sqrt{ g(y')}dy'\right\}^{-1}\right|_{y\to y_0}^{y\to 0^+} -\int_{y_0}^{y\to 0^+} \left\{\frac{q}{g}-\frac{5g'^2}{16g^3}+\frac{g''}{4g^2}\right\}\sqrt{g}dy'.
\eqn
Thus, to calculate the power spectra, one has to perform the integral of $\sqrt{g(y)}$ and the integrals appearing in eq.~(\ref{H}), which are presented in Appendix A. 

\subsection{ Power spectra and circular polarization of PGWs}

With the above general formulas, we can compute the primordial power spectra for each polarization mode of the PGWs as $y \to 0$. The power spectrum for each polarization mode is typically calculated using:
\bqn
\mathcal{P}_{\rm T}^{\rm L} &=& \frac{2 k^3}{\pi^2} \left|\frac{u_{\rm L}(y)}{z_A}\right|^2, \\
\mathcal{P}_{\rm T}^{\rm R} &=& \frac{2 k^3}{\pi^2} \left|\frac{u_{\rm R}(y)}{z_A}\right|^2.
\eqn
Using (\ref{Igg}) and (\ref{errorH}) and after tedious calculations we obtain,
\bqn \label{PS_uniform}
\mathcal{P}_{\rm{T}}^A 
&\simeq &\frac{1}{2}{\cal P}_{\rm T}^{\rm GR, UAA} \left(1 - \sum_{i=1}^{4} d_i \mathcal{C}_i \right),
\eqn
where 
\bqn
{\cal P}_{\rm T}^{\rm GR, UAA} \equiv \frac{362H^2}{9e^3\pi^2}\left[1 + \left(2\ln{2}-\frac{496}{181}\right)\epsilon_1 \right],
\eqn
represents the power spectrum of PGWs calculated from the third-order uniform asymptotic approximation \cite{Zhu:2022dfq, Qiao:2019hkz} and $\mathcal{C}_i$'s represents the contributions from the Lorentz-violating and parity-violating effects, and are given by
\bqn
C_1 &=& \left[1-\frac{10}{543}\beta_\nu (\beta_\nu-1)(\beta_\nu-2)\right]\frac{3^{\beta_{\nu}-1}\sqrt{\pi}\rho_A \Gamma\Big(\frac{\beta_{\nu}}{2}\Big)}{2^{\beta_{\nu}}\Gamma\Big(\frac{\beta_{\nu}+1}{2}\Big)}, \\
C_2 &=& \left[1-\frac{10}{543}\beta_{\bar \nu} (\beta_{\bar \nu}-1)(\beta_{\bar \nu}-2)\right]\frac{3^{\beta_{\bar{\nu}}-1} \sqrt{\pi} \Gamma\Big(\frac{\beta_{\bar{\nu}}}{2}\Big)}{2^{\beta_{\bar{\nu}}}\Gamma\Big(\frac{\beta_{\bar{\nu}}+1}{2}\Big)}, \\
C_3 &=& \left[1-\frac{10}{543}\beta_\mu (\beta_\mu+1)(\beta_\mu+2)\right]\frac{3^{\beta_{\mu}+1} \sqrt{\pi}\rho_A \Gamma\Big(\frac{\beta_{\mu}+2}{2}\Big)}{2^{\beta_{\mu}+2} \Gamma\Big(\frac{\beta_{\mu}+3}{2}\Big)},\\
C_4 &=& \left[1-\frac{10}{543}\beta_{\bar \mu} (\beta_{\bar \mu}+1)(\beta_{\bar \mu}+2)\right]\frac{3^{\beta_{\bar{\mu}}+1} \sqrt{\pi} \Gamma \Big(\frac{\beta_{\bar{\mu}}+2}{2}\Big)}{2^{\beta_{\bar{\mu}}+2} \Gamma\Big(\frac{\beta_{\bar{\mu}}+3}{2}\Big)}.
\eqn
Note that in the above calculations, we have used $v_t \simeq 3/2$ to simplify the expressions of $C_i$'s. Setting ${d}_{1} = {d}_{2} = {d}_{3} = {d}_{4} = 0$ one recovers the standard GR result in the uniform asymptotic approximation. The Lorentz-violating terms ($d_2$ and $d_4$ terms) primarily influence the overall amplitude of the left- and right-handed polarization modes of GW, while the parity-violating terms ($d_1$ and $d_3$ terms) distinctly affect the amplitude of these modes, which tend to amplify the power spectrum of one mode while suppressing that of the other.

We now proceed to evaluate the degree of circular polarization of PGWs, defined as the amplitude difference between the two circular polarization states:
\bqn
\Pi &=& \frac{\mathcal{P}_T^R-\mathcal{P}_T^L}{\mathcal{P}_T^R+\mathcal{P}_T^L} \nb\\
&\simeq& -d_1 \left[1-\frac{10}{543}\beta_\nu (\beta_\nu-1)(\beta_\nu-2)\right] \frac{3^{\beta_{\nu}-1}\sqrt{\pi}\Gamma\left(\frac{\beta_{\nu}}{2}\right)}{2^{\beta_{\nu}}\Gamma\left(\frac{\beta_{\nu}+1}{2}\right)} \nb\\
&& -d_3 \left[1-\frac{10}{543}\beta_\mu (\beta_\mu+1)(\beta_\mu+2)\right]\frac{3^{\beta_{\mu}+1}\sqrt{\pi}\Gamma\left(\frac{\beta_{\mu}+2}{2}\right)}{2^{\beta_{\mu}+2}\Gamma\left(\frac{\beta_{\mu}+3}{2}\right)}. \label{Pi_UAA}
\eqn
As anticipated, the circular polarization degree $\Pi$ is exclusively influenced by parity-violating effects. This formulation aligns with findings from Chern-Simons modified gravity, chiral scalar-tensor theory, and Hořava-Lifshitz gravity. For comprehensive discussions on these theories, we refer to ref.~\cite{Satoh:2008ck}, ref.~\cite{Qiao:2019hkz}, and ref.~\cite{Qiao:2022mln}, respectively.

\section{Some specific examples with exact solutions}
\renewcommand{\theequation}{5.\arabic{equation}}\setcounter{equation}{0}

In this section, we consider several specific cases of $\beta_{\bar \nu}=2$, $\beta_{\bar \mu}=2$, $\beta_{\nu}=1$, and $\beta_{\mu}=-1$, respectively. For these specific cases, the equations of motion of PGWs have exact and analytical solutions. Each case can also arise from one or several specific modified gravities. 

\subsection{$\beta_{\bar{\nu}}=2$}

The case with $\beta_{\bar{\nu}}=2$ can arise from the spatial covariant gravities \cite{Gao:2019liu, Zhu:2022uoq} and the Ho\v{r}ava-Lifshtz gravity with mixed derivative coupling \cite{Colombo:2014lta}. This case corresponds to a specific Lorentz-violating damping rate on the GW propagation. Constraining these effects with currently available data of GW events from LIGO-Virgo-KAGRA collaboration and future ground-based and space-based GW detectors have also been studied in refs. \cite{Zhu:2023wci, Zhang:2024rel} \footnote{In general, the constraints derived from gravitational wave event data analysis cannot be directly applied as bounds on the values of these parameters during the inflationary slow-roll phase. To make this connection requires prior knowledge of how these parameters evolve over time, from the start of slow-roll inflation through to the present era.}. For this case, the equation of motion in eq.~(\ref{u2}) for GWs takes the form
\bqn
u''_A+\left(1+d_2 -\frac{v_t^2-\frac{1}{4}}{k^2\tau^2}\right)k^2 u_A=0.
\eqn
This equation has an exact solution if we treat all the slow-rolling quantities and $d_2$ as constants, which can be expressed analytically as a linear combination of Hankel functions, 
\bqn
u_A \simeq \frac{\sqrt{- \pi \tau}}{2} \Big[c_1 H_{v_t}^{(1)}(- \tilde{k} \tau) + c_2  H_{v_t}^{(2)}(- \tilde{k} \tau)\Big],\label{hankel}
\eqn
where $\tilde{k} =\sqrt{1+d_2} k$, $c_1$ and $c_2$ are two integration constants, and  $H_{v_t}^{(1)}(- \tilde{k} \tau)$ and $ H_{v_t}^{(2)}(- \tilde{k} \tau) $ are the first and second kinds of Hankel function respectively. Using the asymptotic form of the Hankel functions in the limit $- \tilde{k} \tau \to +\infty$ and imposing the normalization condition 
\bqn
\frac{i}{\hbar} (u_A^* u_A' - u^{*}_A{}' u_A) =1, \label{normalization}
\eqn
and initial condition
\bqn\label{initial}
\lim_{y \to +\infty} u_A(y)&\simeq&\frac{1}{\sqrt{2\omega_A}}e^{-i\int\omega_A d\tau},
\eqn
the two constants $c_1$ and $c_2$ can be fixed to be
\bqn
c_1 = 1 , \;\; c_2=0.
\eqn

Then one can use the above exact solution to calculate the power spectra of PGWs in the limit $- \tilde{k} \tau \to 0$. In this limit, the Hankel function $H_{v_t}^{(1)}(- \tilde{k} \tau)$ takes the following asymptotic form
\bqn
\lim_{- \tilde{k} \tau \to 0}H_{v_t}^{(1)}(- \tilde{k} \tau) \simeq  - \frac{i}{\pi}\Gamma(v_t) \left(\frac{-\tilde{k} \tau}{2}\right)^{-v_t},
\eqn
and one finds
\bqn
u_A \to -i \frac{\sqrt{-\pi \tau}}{2\pi} \Gamma(v_t) \left(\frac{-\tilde{k} \tau}{2}\right)^{-v_t}.
\eqn
Then the power spectrum of PGWs reads
\bqn
{\cal P}^{\rm L, R}_{\rm T} \simeq \frac{1}{2} P^{\rm GR}_{\rm T} \left(1- \frac{3}{4} d_2\right), \label{PS_1}
\eqn
where
\bqn
{\cal P}^{\rm GR}_{\rm T} \equiv \frac{2k^2}{\pi^3}\frac{1}{a^2} \Gamma^2(v_t) \left(\frac{- k \tau}{2}\right)^{1-2 v_t}, \label{P_GR}
\eqn
is the power spectrum of PGWs in GR calculated from the exact solution (\ref{hankel}). The Lorentz-violating damping effects with $\beta_\nu=2$ case affects the overall amplitude of the power spectrum of both the left- and right-handed polarization modes of GW in the same way. 

Comparing the above exact result (c.f. (\ref{PS_1})) to that in eq.~(\ref{PS_uniform}), we would like to mention that we have considered the third-order approximation in the uniform asymptotic approximation. The corresponding relative difference between ${\cal P}_{\rm T}^{\rm GR}$ in (\ref{PS_1}) and ${\cal P}_{\rm T}^{\rm GR, UAA}$ in (\ref{PS_uniform}) is less than $0.15\%$ \cite{Zhu:2014aea}. 
Although with such a small error in the calculation of ${\cal P}_{\rm T}^{\rm GR, UAA}$, the small correction to the power spectrum from the Lorentz-violating damping with $\beta_\nu=2$ in the bracket of (\ref{PS_uniform}) exactly reduces to that in the bracket of (\ref{PS_1}).

\subsection{$\beta_{\bar{\mu}}=2$}

The case with $\beta_{\bar{\mu}}=2$ can arise from a large number of Lorentz-violating theories of gravity, including the chiral Weyl gravity \cite{Mylova:2019jrj}, spatial covariant gravities \cite{Gao:2019liu, Zhu:2022uoq}, Ho\v{r}ava-Lifshitz gravity \cite{Horava:2009uw, Blas:2009qj, Zhu:2011xe}, diffeomorphism/Lorentz violating linear gravity in the standard model extension \cite{Mewes:2019dhj, Kostelecky:2017zob}, consistent four-dimensional Einstein-Gauss-Bonnet gravity \cite{Aoki:2020lig, Aoki:2020ila, Li:2022xww}, Lorentz-violating Weyl gravity \cite{Deruelle:2012xv}, etc. This case corresponds to a nonlinear dispersion relation of GWs. The new effect for this case on the PGWs has been studied in ref.~\cite{Aoki:2020ila} with the corresponding exact solution of PGWs. The calculations of the power spectra of PGWs for this case by using the uniform asymptotic approximation has also been carried out in refs. \cite{Ding:2019nwu, Li:2022xww, Zhu:2014aea}. 

For this case, the equation of motion (\ref{u2}) takes the form
\bqn
u''_A+\left(1-\frac{v_t^2-\frac{1}{4}}{k^2\tau^2}+ d_4 k^2\tau^2\right)k^2 u_A=0.
\eqn
By treating all the slow-rolling quantities and $d_4$ as constants, the exact solution of this equation can be written in terms of the Whittaker functions, which is \cite{Aoki:2020ila, Ashoorioon:2011eg}
\bqn
u_A &=& \frac{1}{d_4^{1/4}} \frac{e^{-\frac{\pi}{8\sqrt{d_4}}}}{ \sqrt{-2 \tau} k} \left[c_3 W\left(\frac{i}{4\sqrt{d_4}}, \frac{v_t}{2}, -i \sqrt{d_4}k^2 \tau^2\right) + c_4 M\left(\frac{i}{4\sqrt{d_4}}, \frac{v_t}{2}, - i \sqrt{d_4}k^2 \tau^2\right)\right],\nb\\
\eqn
where $c_3$ and $c_4$ are two integration constants, and $W\left(\frac{i}{4\sqrt{d_4}}, \frac{v_t}{2}, -i \sqrt{d_4}k^2 \tau^2\right)$ and $M\left(\frac{i}{4\sqrt{d_4}}, \frac{v_t}{2}, -i \sqrt{d_4}k^2 \tau^2\right)$ are Whittaker functions \cite{olver_Functions_handbook2010, NIST:DLMF}. Using the asymptotic forms of the Whittaker functions in the limit $- k \tau \to +\infty$ \cite{NIST:DLMF}, i.e.
\bqn
&& W\left(\frac{i}{4\sqrt{d_4}}, \frac{v_t}{2}, -i \sqrt{d_4}k^2 \tau^2\right) \simeq e^{\frac{i}{2}\sqrt{d_4} k^2 \tau^2} (-i \sqrt{d_4} k^2 \tau^2)^{\frac{i}{4\sqrt{d_4}}}, \\
&& M\left(\frac{i}{4\sqrt{d_4}}, \frac{v_t}{2}, -i \sqrt{d_4}k^2 \tau^2\right) \simeq \frac{\Gamma(1+v_t)}{\Gamma\left(\frac{1+v_t}{2}-\frac{i}{4\sqrt{d_4}}\right)}e^{-\frac{i}{2}\sqrt{d_4} k^2 \tau^2} (-i \sqrt{d_4} k^2 \tau^2)^{-\frac{i}{4\sqrt{d_4}}},\nb\\
\eqn
and imposing the normalization condition (\ref{normalization}) and initial condition (\ref{initial}), the two constants $c_3$ and $c_4$ can be fixed to be
\bqn
c_3 = 1 , \;\; c_4=0.
\eqn

Then one can use the above exact solution to calculate the power spectra of PGWs in the limit $- k \tau \to 0$. In this limit, the Whittaker function $W\left(\frac{i}{4\sqrt{d_4}}, \frac{v_t}{2}, -i \sqrt{d_4}k^2 \tau^2\right)$ takes the following asymptotic form
\bqn
&&\lim_{- k \tau \to 0} W\left(\frac{i}{4\sqrt{d_4}}, \frac{v_t}{2}, -i \sqrt{d_4}k^2 \tau^2\right) \simeq  \frac{\left(-i \sqrt{d_4} k^2 \tau^2\right)^{\frac{1}{2}-\frac{v_t}{2}} \Gamma(v_t)}{\Gamma\left(\frac{1}{2}+\frac{v_t}{2}-\frac{i}{4\sqrt{d_4}}\right)},
\eqn
and one finds
\bqn
u_A \to \frac{1}{\sqrt{2k}} (-i)^{\frac{1-v_t}{2}} e^{- \frac{\pi}{8\sqrt{d_4}}} (- k\tau)^{\frac{1}{2}-v_t}.
\eqn
Then the power spectrum of PGWs reads
\bqn
{\cal P}^{\rm L, R}_{\rm T} = \frac{1}{2} P^{\rm GR}_{\rm T} \frac{ 2^{2 v_t} \pi d_4^{-\frac{v_t}{2}}e^{- \frac{\pi}{4\sqrt{d_4}}}}{\left|\Gamma\left(\frac{1}{2}+\frac{v_t}{2}-\frac{i}{4\sqrt{d_4}}\right)\right|^2}, \label{PS_2}
\eqn
where ${\cal P}^{\rm GR}_{\rm T} $ is given by eq.~(\ref{P_GR}). Considering the correction term $d_4$ is small, one can employ the following asymptotic form of Gamma function $\Gamma(x+i y)$ (note that $x$ and $y$ are both real) for large $y$ \cite{Zhu:2017jew} \footnote{see appendix B in \cite{Zhu:2017jew} for the derivation of this asymptotic form.}, 
\bqn
&& |\Gamma(x+iy)| \simeq \sqrt{2\pi} |y|^{x- \frac{1}{2}} e^{- \frac{\pi y}{2}}  \times \left[1+ \left(\frac{x^3}{6}-\frac{x^2}{4}+\frac{x}{12}\right) \frac{1}{y^2} + \mathcal{O}\left(\frac{1}{y^4}\right)\right],
\eqn
and then the power spectrum of PGWs can be approximately expressed as
\bqn
{\cal P}^{\rm L, R}_{\rm T} \simeq \frac{1}{2} P^{\rm GR}_{\rm T} \left(1 - \frac{5}{4} d_4\right). \label{PS_22}
\eqn
Similar to the $\beta_\nu=2$ case, the Lorentz violation with $\beta_{\bar \mu}=2$ case only affects the overall amplitude of the power spectrum of PGWs. 

Comparing to (\ref{PS_uniform}), as we already mentioned in the last subsection, the corresponding relative difference between ${\cal P}_{\rm T}^{\rm GR}$ in (\ref{PS_1}) and ${\cal P}_{\rm T}^{\rm GR, UAA}$ in (\ref{PS_uniform}) is less than $0.15\%$ \cite{Zhu:2014aea}. We also observe a $ \sim 0.4\%$ relative difference in the coefficient of $d_4$ in the brackets of (\ref{PS_22}) and (\ref{PS_uniform}), while the value of the coefficient in (\ref{PS_uniform}) is $-\frac{909}{724}$ for $\beta_{\bar \mu}=2$ case.

\subsection{$\beta_\nu=1$ and $\beta_{\mu}=-1$}

The cases for $\beta_\nu=1$ and $\beta_{\mu}=-1$ share the same form in equation eq.~(\ref{u2}), thus we discuss them together in this subsection. These two cases correspond to two distinct effects. The $\beta_\nu=1$ case corresponds to the amplitude birefringence of GWs, resulting in different and frequency-dependent damping rates of left- and right-hand circular polarizations of GWs. This case can arise from Chern-Simons gravity \cite{Jackiw:2003pm, Alexander:2009tp, Alexander:2004wk}, Palatini-Chern-Simons gravity \cite{Sulantay:2022sag}, spatial covariant gravities \cite{Gao:2019liu, Zhu:2022uoq}, chiral scalar-tensor theory \cite{Crisostomi:2017ugk, Qiao:2019hkz}, parity-violating scalar-nonmetricity theory \cite{Chen:2022wtz, Conroy:2019ibo, Li:2022vtn}, and a parity-violating coupling between a  Kalb-Ramond field and the Riemann curvature \cite{Manton:2024hyc}. The studies of PGWs for this case can be found in refs.~\cite{Satoh:2008ck, Alexander:2004wk}. The $\beta_{\mu}=-1$ case corresponds to velocity birefringence, leading to different and frequency-dependent velocities of left- and right-hand circular polarizations of GWs. This case can appear in parity-violating scalar-nonmetricity theory \cite{Chen:2022wtz, Conroy:2019ibo, Li:2022vtn}, metric-affine Chern-Simons gravity \cite{Boudet:2022nub}, Nieh-Yan teleparallel modified gravity \cite{Li:2020xjt, Li:2021wij, Wu:2021ndf}, diffeomorphism-violating linear gravity in the standard model extension \cite{Kostelecky:2017zob}, the new GR with parity violation \cite{Hohmann:2022wrk}, and a parity-violating coupling between a  Kalb-Ramond field and the Riemann curvature \cite{Manton:2024hyc}. The polarized PGWs for this case have been briefly discussed in parity-violating scalar-nonmetricity theory in \cite{Chen:2022wtz}.

For these two cases, the equation of motion (\ref{u2}) takes the form
\bqn
u''_A+\left(1-\frac{v_t^2-\frac{1}{4}}{k^2\tau^2}+ \frac{d_{13}^A}{k \tau} \right)k^2 u_A=0,
\eqn
where $d_{13}^A \equiv \rho_A (d_1+d_3)$. By treating all the slow-rolling quantities and $d_{13}^  A$ as constants, the exact solution of this equation can be written in terms of the Whittaker functions \cite{olver_Functions_handbook2010, NIST:DLMF},
\bqn
u_A &=& \frac{(2ik\tau)^{\frac{i d_{13}^A}{2}}}{\sqrt{2k}} \left[c_5 W\left(-\frac{i d_{13}^A}{2}, v_t, 2i k \tau\right) + c_6 M\left(-\frac{i d_{13}^A}{2}, v_t, 2i k \tau\right)\right],
\eqn
where $c_5$ and $c_6$ are two integration constants. Using the asymptotic forms of the Whittaker functions in the limit $- k \tau \to +\infty$ \cite{NIST:DLMF}, i.e.,
\bqn
&& W\left(-\frac{i d_{13}^A}{2}, v_t, 2i k \tau\right) \simeq e^{- i k \tau} (2i k \tau)^{-\frac{i d_{13}^A}{2}}, \\
&& M\left(-\frac{i d_{13}^A}{2}, v_t, 2i k \tau\right) \simeq \frac{\Gamma(1+2v_t)}{\Gamma\left(\frac{1}{2}+v_t-\frac{i d_{13}^A}{2}\right)} e^{i k \tau} (2i k \tau)^{\frac{i d_{13}^A}{2}},
\eqn
and imposing the normalization condition in eq.~(\ref{normalization}) and initial condition (\ref{initial}), the two constants $c_5$ and $c_6$ can be fixed to be
\bqn
c_5 = 1 , \;\; c_6=0.
\eqn

Then one can use the above exact solution to calculate the power spectra of PGWs in the limit $- k \tau \to 0$. In this limit, the Whittaker function $W\left(-\frac{i d_{13}^A}{2}, v_t, 2i k \tau\right)$ takes the following asymptotic form
\bqn
&&\lim_{- k \tau \to 0} W\left(-\frac{i d_{13}^A}{2}, v_t, 2i k \tau\right) \simeq  \frac{\left(2i k \tau\right)^{\frac{1}{2}-v_t} \Gamma(2v_t)}{\Gamma\left(\frac{1}{2}+v_t+\frac{i d_{13}^A}{2}\right)}.
\eqn
Then the power spectrum of PGWs reads
\bqn
{\cal P}^A_{\rm T} = \frac{1}{2} P^{\rm GR}_{\rm T} \frac{e^{- \frac{\pi d_{13}^A}{2}}\left|\Gamma\left(\frac{1}{2}+v_t\right)\right|^2}{\left|\Gamma\left(\frac{1}{2}+v_t+\frac{i d_{13}^A}{2}\right)\right|^2},\label{PS_3}
\eqn
where ${\cal P}^{\rm GR}_{\rm T} $ is given by eq.~(\ref{P_GR}). Considering that the correction term $d_{13}^A$ is small, the power spectrum  can be approximately expressed as
\bqn
{\cal P}^A_{\rm T} \simeq \frac{1}{2} P^{\rm GR}_{\rm T} \left[1 - \frac{\pi}{2} \rho_A (d_1+d_3)\right]. \label{PS_33}
\eqn
Comparing the above result to (\ref{PS_uniform}), it is not difficult to check that the correction term in the bracket of (\ref{PS_uniform}) exactly reduces to that in (\ref{PS_33}). The cases for $\beta_\nu=1$ and $\beta_\mu=-1$ affect the power spectra of left- and right-handed polarization modes of PGWs in different ways. This effect is due to the violation of the parity symmetry in gravity. For a positive value of $d_1+d_3$ in the above expression, the new effect trends to enhance (suppress) the power spectra of the left (right) -handed modes. 

Because of the difference between the two modes of PGWs, the circular polarization of PGWs can be expressed as
\bqn
\Pi \simeq - \frac{\pi}{2}(d_1+d_3),
\eqn
which is the same as the result given in (\ref{Pi_UAA}) for cases with $\beta_\nu=1$ and $\beta_{\mu}=-1$ from the uniform asymptotic approximation.

\section{Conclusions}
\renewcommand{\theequation}{6.\arabic{equation}}\setcounter{equation}{0}

Violations of parity and Lorentz symmetries in gravity can result in several distinct effects on the propagation of GWs. The parity violation in general leads to an asymmetry of the propagation speed and amplitude damping between left- and right-hand polarizations of GW, which induces the velocity and amplitude birefringence, respectively. Lorentz violation, on the other hand, can result in two distinct effects: one is the modified dispersion relation and another is a frequency-dependent damping of GWs. These new effects can arise from a large number of parity- and Lorentz-violating theories of gravity. 

In this paper, to study the parity- and Lorentz-violating effects on both power spectra and circular polarization of PGWs, we employ a systematic parametric framework constructed in refs.~\cite{Zhao:2019xmm, Zhu:2023wci}, for characterizing possible derivations of GW propagation in GR. It is also shown in refs.~\cite{Zhao:2019xmm, Zhu:2023wci} that the GW propagations in a large number of parity- and Lorentz-violating theories of gravity can be well described by this parameterization. 

With this parameterization, we calculate explicitly both the power spectra and for the two polarization modes and the corresponding degree of circular polarization of PGWs with parity- and Lorentz-violating corrections in a unified way. To verify the validity and accuracy of our approximate results, we also compare our approximate power spectra and the circular polarization derived from the uniform asymptotic approximation with several special cases that have exact results. It is shown that the approximate results from the uniform asymptotic approximation fit the exact results extremely well. 

Our results show that the power spectra can be modified due to the presence of both parity and Lorentz violations. The Lorentz violation only affects the overall amplitude of both left- and right-handed polarization modes of GW. The parity violation, on the other hand, affects the amplitudes of left- and right-handed polarization modes of GW in different ways. Because of this, the degree of circular polarization becomes nonzero, which is directly related to parity-violating parameters $d_1$ and $d_3$. However, this effect is expected to be very small due to the suppression of the parity-violating parameters $d_1, d_3 < \mathcal{O}(1)$ and it seems very difficult to detect it by using the power spectra of future CMB data, as shown in detail in ref.~ \cite{Wang:2012fi}. 

Here we would like to mention that in our study we only consider the power spectra and the corresponding circular polarization of PGWs. It is pointed out in \cite{Bartolo:2017szm} that the parity-violating effects in the non-Gaussianities of PGWs could be large enough and detectable in the future CMB data and survey of the large-scale structure of the Universe. Thus it is interesting to explore the possible signatures of the circular polarization of PGWs in non-Gaussianities, large-scale structure, and EB correlation in the galaxy-shaped power spectrum, etc. We would like to come back to these issues in our future works.

\acknowledgments

This work is supported in part by the National Key Research and Development Program of China under Grant No.2020YFC2201503, the National Natural Science Foundation of China under Grants No.12275238 and No. 11675143, the Zhejiang Provincial Natural Science Foundation of China under Grants No.LR21A050001 and No. LY20A050002, and the Fundamental Research Funds for the Provincial Universities of Zhejiang in China under Grant No. RF-A2019015. WZ is supported by the National Key R\&D Program of China Grant No. 2021YFC2203102 and 2022YFC2204602, Strategic Priority Research Program of the Chinese Academy of Science Grant No. XDB0550300, NSFC No. 12325301 and 12273035, the Fundamental Research Funds for the Central Universities under Grant No. WK3440000004.


\appendix

\section{Calculations of the integral of $\sqrt{g(y)}$ and the error control function $\mathscr{H}(+\infty)$}
\renewcommand{\theequation}{A.\arabic{equation}}\setcounter{equation}{0}

In this appendix, we present the detailed calculation of the integral of $\sqrt{g(y)}$ in eq.~(\ref{PW_3}) and the error control function $\mathscr{H}(+\infty)$ in eq.~(\ref{H}). To perform these integrals, our strategy here is to treat all the new terms that arise from the parity and Lorentz violations as small corrections. With this consideration, we can expand the integrals in terms of the small coefficients $d_1$, $d_2$, $d_3$, and $d_4$. Most of the formulas or expressions used here can also be found in the Appendix B.2 in ref.~\cite{Zhu:2015ata}.

To expand an integral in terms of a small parameter $\epsilon$, let us consider the following formula
\bqn
I[a(\epsilon),b(\epsilon),\epsilon]=\int_{a(\epsilon)}^{b(\epsilon)} F[y',\epsilon]dy'.
\eqn
Now expanding the above integral in terms of $\epsilon$ yields
\bqn\lb{derint}
I[a(\epsilon),b(\epsilon),\epsilon]\simeq \int_{a(0)}^{b(0)} F(y',0)dy'+\epsilon \int_{a(0)}^{b(0)} F_{,\epsilon}(y',0)dy'+\epsilon \Big[b_{,\epsilon}(0)F(b_0,0)-a_{,\epsilon}(0)F(a_0,0)\Big].\nb\\
\eqn
Here ``," in $F_{, \epsilon}$, $a_{, \epsilon}$, and $b_{, \epsilon}$ denotes the derivative with respect to the small parameter $\epsilon$. Using this formula, the integral of $\sqrt{g(y)}$ in eq.~(\ref{PW_3}) can be expanded in terms of the coefficients $d_i\; (i=1, 2, 3, 4)$ as
\bqn\lb{int_of_sqrt_g}
\int^{y_0}_{y}\sqrt{g(y')}dy' &\simeq & \int^{v_t}_{y} \sqrt{g(y')}\Big|_{d_1, d_2, d_3, d_4=0}dy'\nb\\
&& + \sum_{i=1}^4 d_i \Bigg\{\int_{y}^{v_t} \left.\frac{g_{,d_i}(y')}{2\sqrt{g(y')}}\right|_{d_1, d_2, d_3, d_4=0}dy'  - y_{0, d_i} \sqrt{g(v_t)}\Big|_{d_1, d_2, d_3, d_4=0}\Bigg\}.\nb\\
\eqn
Note that in the above expression, we have $g(v_t)|_{d_1, d_2, d_3, d_4}=0$. Directly performing the above two integrals, one obtains
\bqn
 \lim_{y \to 0^+}\int^{y_0}_{y}\sqrt{g(y')}dy' &\simeq &- \left(1+ \ln \frac{y}{2v_t}\right)v_t   - d_1 \frac{v_t^{\beta_{\nu}-1} \rho_A  \sqrt{\pi} \Gamma(\frac{\beta_\nu}{2})}{4 \Gamma(\frac{1+\beta_{\nu}}{2})} - d_2 \frac{v_t^{\beta_{\bar \nu}-1} \sqrt{\pi} \Gamma(\frac{\beta_{\bar \nu}}{2})}{4 \Gamma(\frac{\beta_{\bar \nu}+1}{2})} \nb\\
 &&- d_3 \frac{v_t^{\beta_\mu+1} \rho_A \sqrt{\pi} \Gamma(1+\frac{\beta_\mu}{2})}{4 \Gamma(\frac{3+\beta_\mu}{2})}  - d_4 \frac{v_t^{\beta_{\bar \mu}+1} \rho_A \sqrt{\pi} \Gamma(1+\frac{\beta_{\bar\mu}}{2})}{4 \Gamma(\frac{3+\beta_{\bar \mu}}{2})}. \label{Igg}
\eqn

Now let us turn to the error control function $\mathscr{H}(+\infty)$ in eq.~(\ref{H}), from which we can divide $\mathscr{H}(+\infty)$ into two parts. We first consider the first part, which is
\bqn
&&\frac{5}{36}\left.\frac{1}{I_1[y,y_0,d_i]}\right|_{y_0-\varepsilon}^{0^+}=\frac{5}{36}\left\{\frac{1}{I_1[0,y_0,d_i]}-\frac{1}{I_1[ y_0-\varepsilon,y_0,d_i]}\right\},
\eqn
where
\bqn\lb{B5}
I_1[0, y_0,d_i]&=&\lim_{y\to 0^+}\int_{ y_0}^{y}\sqrt{g(y')}dy',\nb\\
I_1[y_0-\varepsilon,y_0,d_i]&=&\int_{y_0}^{y_0-\varepsilon }\sqrt{g(y')}dy'.
\eqn
Here $\varepsilon$ is a positive and small quantity, representing the divergences in the expressions. The integral of $I_1[0, y_0,d_i]$ is already calculated in eq.~(\ref{Igg}), which contains a divergent term $\ln \frac{y}{2 v_t}$ in the limit $y\to 0^+$. Thus we have
\bqn\lb{I0}
\frac{5}{36}\frac{1}{I_1[0,y_0,d_i]}=0.
\eqn
Similarly, for $I_1[y_0-\varepsilon,y_0,d_i]$, using (\ref{derint}) one has
\bqn
I_1[y_0-\varepsilon,y_0, d_i]&\simeq&\lim_{\varepsilon\to 0} \int_{v_t}^{v_t-\varepsilon}\sqrt{g(y')}\Big|_{d_1, d_2, d_3, d_4=0}dy' + \sum_{i=1}^4 d_i\lim_{\varepsilon \to 0} I_{1,d_i}[v_t-\varepsilon,v_t,0],\nb\\
\eqn
where
\bqn\lb{B9}
I_{1,d_i}[v_t-\varepsilon,v_t,0] &=& \int_{v_t}^{v_t-\epsilon} \frac{g_{,d_i}(y')}{2\sqrt{g(y')}}\Big|_{d_1, d_2, d_3, d_4=0}dy'\nb\\
&& + y_{0,d_i} \sqrt{g(v_t-\varepsilon)}\Big|_{d_1, d_2, d_3, d_4=0}  -y_{ 0,d_i} \sqrt{g(v_t)}\Big|_{d_1, d_2, d_3, d_4=0}.
\eqn
Using this expression we find
\bqn
\frac{5}{36I_1[y_0-\varepsilon,y_0,d_i]}  \simeq  \frac{5}{36I_1[v_t-\varepsilon,v_t,0]} - \sum_{i=1}^4 d_i\frac{5I_{1,d_i}[v_t-\varepsilon,v_t,0]}{36I_1[v_t-\varepsilon,v_t,0]^2 }.
\eqn
Then combined with eq.~(\ref{I0}), the first part of the error control function can be calculated by using the following formula
\bqn
\frac{5}{36}\left.\frac{1}{I_1[y,y_0,d_i]}\right|_{y=y_0-\varepsilon}^{y=0^+}\simeq -\frac{5}{36I_1[v_t-\varepsilon,v_t,0]}+\sum_{i=1}^4d_i\frac{5I_{1,d_i}[v_t-\varepsilon,v_t,0]}{36I_1[v_t-\varepsilon,v_t,0]^2 }.
\eqn

Now let us turn to consider the second part of the error control function. Let us first define
\bqn
I_2[0,y_0-\varepsilon,d_i] \equiv -\int_{y_0-\varepsilon}^{0^+} \left\{\frac{q}{g}-\frac{5g'^2}{16g^3}+\frac{g''}{4g^2}\right\}\sqrt{g}dy'=\int_{y_0-\varepsilon}^{0^+} G(y')dy',
\eqn
with
\bqn
G(y)\equiv-\left\{\frac{q}{g}-\frac{5g'^2}{16g^3}+\frac{g''}{4g^2}\right\}\sqrt{g}.
\eqn
According to the formula (\ref{derint}) we find
\bqn\lb{B14}
I_2[0,y_0-\varepsilon,d_i] &\simeq& \int_{v_t-\varepsilon}^{0^+} G(y')\Big|_{d_1, d_2, d_3, d_4=0}dy'\nb\\
&&+\sum_{i=1}^4 d_i \int_{v_t-\varepsilon}^{0^+} G_{,d_i}(y')\Big|_{d_1, d_2, d_3, d_4=0}dy' - \sum_{i=1}^4 d_i y_{0,d_i} G(v_t-\varepsilon)\Big|_{d_1, d_2, d_3, d_4=0}.\nb\\
\eqn
Thus finally we can calculate the error control function by using the following formulas
\bqn\lb{int_exp}
\mathscr{H}(+\infty) &\simeq& -\frac{5}{36I_1[v_t-\varepsilon,v_t,0]}+ \sum_{i=1}^4d_i\frac{5I_{1,d_i}[v_t-\varepsilon,v_t,0]}{36I_1[v_t-\varepsilon,v_t,0]^2 }+I_2[0,y_0-\varepsilon,d_i],
\eqn
where $I_1[v_t-\varepsilon,v_t,0]$, $I_{1,d_i}[v_t-\varepsilon,v_t,0]$, and $I_2[0,y_0-\varepsilon,d_i]$ are given by eq.~(\ref{B5}), eq.~(\ref{B9}), and eq.~(\ref{B14}), respectively. Then performing the integrals in the expressions of $I_1[v_t-\varepsilon,v_t,0]$, $I_{1,d_i}[v_t-\varepsilon,v_t,0]$, and $I_2[0,y_0-\varepsilon,d_i]$, and after tedious calculations, one obtains
\bqn\label{errorH}
\mathscr{H}(+\infty) &\simeq&  \frac{1}{6 v_t} + d_1 \frac{\beta_\nu (\beta_\nu-1)(\beta_\nu-2) }{24 v_t^2}\frac{v_t^{\beta_{\nu}-1}\sqrt{\pi}\rho_A \Gamma\Big(\frac{\beta_{\nu}}{2}\Big)}{2\Gamma\Big(\frac{\beta_{\nu}+1}{2}\Big)} \nb\\
&& +d_2 \frac{\beta_{\bar \nu} (\beta_{\bar \nu}-1)(\beta_{\bar \nu}-2) }{24 v_t^2}\frac{v_t^{\beta_{\bar \nu}-1}\sqrt{\pi}\rho_A \Gamma\Big(\frac{\beta_{\bar \nu}}{2}\Big)}{2\Gamma\Big(\frac{\beta_{\bar \nu}+1}{2}\Big)} \nb\\
&&+ d_3 \frac{\beta_\mu (\beta_\mu +1)(\beta_\mu+2) }{24 v_t^2}\frac{v_t^{\beta_{\mu}+1}\sqrt{\pi}\rho_A \Gamma\Big(\frac{\beta_{\mu}+2}{2}\Big)}{2\Gamma\Big(\frac{\beta_{\mu}+3}{2}\Big)} \nb\\
&& + d_4 \frac{\beta_{\bar \mu} (\beta_{\bar \mu} +1)(\beta_{\bar \mu}+2) }{24 v_t^2}\frac{v_t^{\beta_{\bar \mu}+1}\sqrt{\pi} \Gamma\Big(\frac{\beta_{\bar \mu}+2}{2}\Big)}{2\Gamma\Big(\frac{\beta_{\bar \mu}+3}{2}\Big)}.
\eqn





\end{document}